\documentclass[10pt,prl,aps,twocolumn,showpacs,floatfix]{revtex4}
\usepackage{graphicx}

\usepackage{amssymb}
\usepackage{amsmath}

\begin{document}

\title{Method to characterize spinons as emergent elementary particles}

\author{Ying Tang and Anders W. Sandvik}
\affiliation{Department of Physics, Boston University, 590 Commonwealth Avenue, Boston, Massachusetts 02215}

\begin{abstract}
We develop a technique to directly study spinons (emergent spin $S=1/2$ particles) in quantum spin
models in any number of dimensions. The size of a spinon wave packet and of a bound pair (a triplon)
are defined in terms of wave-function overlaps that can be evaluated by quantum Monte Carlo simulations. 
We show that the same information is contained in the spin-spin correlation function as well. We illustrate 
the method in one dimension. We confirm that spinons are well defined particles (have exponentially 
localized wave packet) in a valence-bond-solid state, are marginally defined (with power-law shaped 
wave packet) in the standard Heisenberg critical state, and are not well defined in an ordered N\'eel 
state (achieved in one dimension using long-range interactions).
\end{abstract}

\date{\today}


\maketitle

Spinons are emergent spin $S=1/2$ particles (fractional excitations) of quantum magnets 
\cite{shastry81,faddeev81,wen04} and potentially exist also in strongly-correlated electron systems such 
as the high-T$_{\rm c}$ cuprate superconductors \cite{anderson96}. Their existence is well established 
in one-dimensional (1D) systems \cite{shastry81,faddeev81}, where they correspond to kinks and antikinks 
(solitons). In higher dimensions, gapped magnons (``triplons'') can be viewed as bound states of spinons. 
Under some conditions, in spin liquid states \cite{wen04} and at certain quantum-critical points 
\cite{senthil04}, these spinons may become deconfined (unbound).
Even in cases where the spinons are not completely deconfined, such as in a valence-bond-solid (VBS) 
state of a two-dimensional (2D) system close to a phase transition into the antiferromagnetic (N\'eel) state, 
the bound state can become very large \cite{senthil04}. The spinons can then be viewed as deconfined below the 
length-scale of the pair size, and above a corresponding (relatively low) energy scale. This is analogous 
to quarks, which are the elementary constituent particles of the baryons although they are 
strictly speaking always confined.

Observing deconfined or almost deconfined spinons in experiments is in general difficult \cite{zhou11}. In 1D 
systems, e.g., the Heisenberg chain, it is well understood (based on the exact Bethe ansatz solution and
numerical calculations \cite{faddeev81,pereira06}) that spinons lead to a broad continuum in the dynamic spin 
structure factor $S(q,\omega)$. 
%due to pairs of spinons with momenta $q_1$ and $q_2$ with $q=q_1+q_2$. 
This continuum has been observed in neutron scattering experiments on quasi-1D quantum antiferromagnets \cite{tennant93}. In 2D 
systems, there is no known reference model with deconfined spinons in which $S(q,\omega)$ can be computed exactly. One 
nevertheless expects a broad continuum also in this case, and such experimental signatures have been 
claimed in some quasi-2D systems \cite{headings10}. The issue is complicated, however, by the fact that 
a continuum is also expected due to multi-magnon processes \cite{sandvik01}.

In this {\it Letter}, we discuss spinon detection in numerical model calculations. This has also been a 
challenging problem, the solution of which will greatly help to understand the conditions under which 
spinons can exist as independent elementary particles. Recently, signatures in thermodynamic properties 
were observed \cite{sandvik11} in a 2D J-Q model (a spin-$1/2$ Heisenberg model including  
four-spin interactions \cite{sandvik07}) at the point separating its N\'eel and VBS ground
states. This model may, thus, exhibit the deconfined quantum-criticality proposed by Senthil 
{\it et al.}~\cite{senthil04}. It is still desirable to have a more direct way to unambiguously (independently
of any phenomenological ansatz or theory) detect spinons in numerical studies of spin models 
(and eventually in doped systems). Here we introduce a method based on quantum Monte 
Carlo (QMC) simulations in the basis of valence bonds (singlet pairs) \cite{sandvik05}, generalized to 
include one or two unpaired spins \cite{wang10,banerjee10}. We show that an unpaired spin can constitute
the core of a spinon wave packet, the size of which can be computed with our method. 
Analyzing the separation of two such wave packets we obtain quantitative information 
on the confinement or deconfinement of spinons in a magnetically disordered state. 
Importantly, our definitions also reproduce the expectation that the
spinon should not be a low-energy particle in the ordered N\'eel state.

{\it Models}---The primary model we use to test our method is the J-Q$_{\rm 3}$ chain, a 1D member 
of the broad class of J-Q models introduced in Refs.~\cite{sandvik07,lou09}. Defining a two-spin singlet
projector $C_{i,j}=1/4-{\bf S}_i\cdot {\bf S}_j$, the hamiltonian is
\begin{equation}
H = -\sum_{i=1}^N (JC_{i,i+1}+Q_3C_{i,i+1}C_{i+2,i+3}C_{i+4,i+5}),
\label{jqham}
\end{equation}
where $J,Q_3\ge 0$ and we define $g=Q_3/J$. The ground state of this system is in the class of the 
standard critical Heisenberg chain for $g<g_c$ and is a doubly-degenerate VBS for $g>g_c$. Using Lanczos 
diagonalization to extract the lowest singlet and triplet excitations and studying their crossings in the 
standard way for this kind of transition (see, e.g., \cite{sandvik10b}), we obtain $g_c\approx 0.1645$ 
(in agreement with a recent QMC study of the critical properties of the same model \cite{kedar11}). We 
have also studied the J-Q$_{\rm 2}$ model, i.e., using two singlet projectors in the Q-term in (\ref{jqham}),
for which $g_c\approx 0.84831$. We here focus on the J-Q$_{\rm 3}$ model because it is more strongly 
VBS ordered at $J=0$.

\begin{figure}
\centerline{\includegraphics[width=6.75cm, clip]{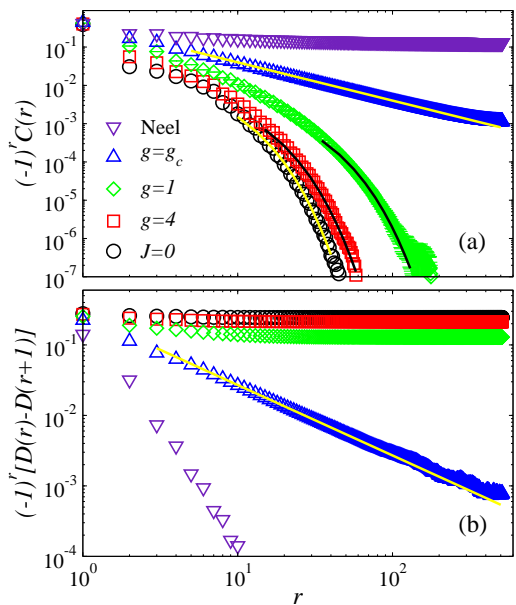}}
\vskip-2mm
\caption{(Color online) Spin (a) and dimer (b) correlations of systems with $N=1024$ spins. Results 
for the J-Q$_{\rm 3}$ model in the VBS phase ($J=0$, $g=4,1$) and at criticality ($g_c$) are shown 
along with the behavior in the N\'eel state of the long-range model with $\alpha=3/2$. The curves in (a) 
are fits to the form  $\propto {\rm e}^{-r/\xi}$ (with $\xi \approx 4$ at $J=0$). The straight lines 
at the $g_c$ data show the expected $\sim 1/r$ critical behavior \cite{eggert96}.} 
\label{fig1}
\vskip-3mm
\end{figure}

We also wish to study an ordered N\'eel state, which in an SU(2) invariant 1D system can only be 
achieved with long-range interactions. The hamiltonian
\begin{equation}
H = \sum_{i=1}^N \sum_{r=1}^{N/2}(-1)^{r-1} J_r {\bf S}_i\cdot {\bf S}_{i+r},~~~J_r>0,
\end{equation}
was studied in \cite{laflorencie05}. With $J_r=1/r^\alpha$, a quantum phase transition from the critical state 
for $\alpha>\alpha_c$ to a N\'eel state for $\alpha<\alpha_c$ was observed, with $\alpha_c \approx 2.2$. Here we use 
a slightly different model, with $J_r=1/r^\alpha$ for odd $r$ but $J_r=0$ for even $r$, to make the system amenable to 
QMC simulations in the valence-bond basis \cite{sandvik05}. We choose $\alpha=3/2$, for which the system is N\'eel ordered.

To demonstrate the ground states of interest---VBS, critical, and N\'eel---in Fig.~\ref{fig1} 
we plot the spin and dimer correlation functions, defined by
\begin{align}
C(r)&=\langle {\bf S}_i\cdot {\bf S}_{i+r}\rangle, \label{cr} \\
D(r)&=\langle ({\bf S}_i\cdot {\bf S}_{i+1})({\bf S}_{i+r}\cdot {\bf S}_{i+1+r})\rangle,\label{dr}
\end{align}
and computed using the QMC method discussed below. We multiply $C(r)$ by $(-1)^r$ to cancel the 
signs of the correlations and graph $(-1)^r[D(r)-D(r+1)]$, which for large $r$ can be regarded 
as the VBS order parameter.

{\it QMC method}---The valence-bond QMC algorithm and its generalizations to $S>0$ states have been
discussed in several papers \cite{sandvik05,wang10,sandvik10c,banerjee10}. Here we review key aspects 
of the basis and the form of the generated ground states.

Acting with a high power of the hamiltonian $H^m$ on a trial state $|\Psi_t\rangle$, with $H$ written as 
a sum of singlet projectors (individual ones and products of three, for J and Q interactions,
respectively), the ground-state normalization $\langle \Psi_0|\Psi_0\rangle$ is sampled (for $m$ 
large enough for $H^m|\Psi_t\rangle$ to be completely dominated by $|\Psi_0\rangle$).
In an $S=0$ state for even $N$, the states are expressed as superpositions 
of bipartite valence-bond states $|V_\alpha\rangle$, i.e., products of $N/2$ singlets $(a,b)=
(\uparrow_a\downarrow_b - \downarrow_b\uparrow_a)/\sqrt{2}$ where $a$ and $b$ are sites on sublattice 
$A$ and $B$, respectively. We use trial states of the amplitude-product form \cite{liang88}.

\begin{figure}
\centerline{\includegraphics[width=7.25cm, clip]{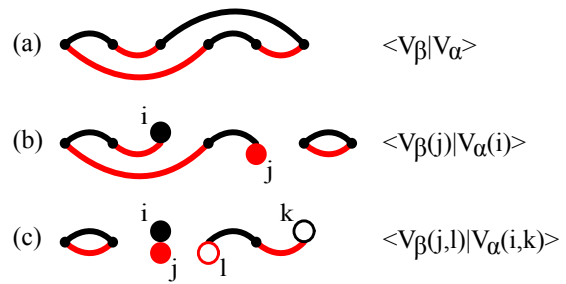}}
\vskip-2mm
\caption{(Color online) Illustration of the basis for states with (a) $S=0$ (even $N$), (b) $S=1/2$ 
(odd $N$), and (c) $S=1$ (even $N$). The bonds and unpaired spins of the bra and ket states are shown 
below and above the line of sites, respectively.} 
\label{fig2}
\vskip-3mm
\end{figure}

The valence-bond basis is non-orthogonal, and the normalization of the projected ground state is therefore
of the form $\langle \Psi_0|\Psi_0\rangle=\sum_{\alpha\beta}f_\beta f_\alpha \langle V_\beta|V_\alpha \rangle$,
where $f_\beta,f_\alpha$ are not known explicitly. Implicitly, the probability of generating a pair of states
is $P(V_\alpha,V_\beta)=f_\beta f_\alpha \langle V_\beta|V_\alpha \rangle$. The overlap 
$\langle V_\beta|V_\alpha \rangle = 2^{N_\circ-N/2}$, where $N_\circ$ is the number of loops 
in the transition graph of the two states. Fig.~\ref{fig2}(a) shows a case with $N_\circ=1$. Matrix 
elements of the form $\langle V_\beta|A|V_\alpha \rangle$ for many observables $A$ of interest depend 
on the loop structure of the transition graph \cite{liang88,beach05}.

For $S>0$ and magnetization $m_z=S$ the states have $2m_z$ unpaired $\uparrow$ spins and $(N-2m_z)/2$ singlet 
bonds (as discussed, e.g., in \cite{wang10,banerjee10}). For odd $N$, which we use for $S=1/2$, the system is in 
principle frustrated by periodic boundaries. This is a finite-size effect, however, 
which vanishes when $N\to \infty$ (at least for observables probing distances $r \ll N$). The QMC loop updates \cite{sandvik10c} automatically exclude frustrated 
negative-sign configurations, and this should, thus, be the most rapid way to approach $N=\infty$. Configurations 
for $S=1/2$ and  $S=1$ states are illustrated in Fig.~\ref{fig2}(b,c). We note that the valence bond basis 
with two unpaired spins was used in a pioneering variational study on spinon deconfinement in a VBS state
of a 1D frustrated model \cite{shastry81}. 

{\it Spinon statistics}---
The first aspect of our method relies on the representation of $S=1/2$ states in terms of valence-bond states with 
an unpaired spin \cite{banerjee10}. One can determine whether there is a well-defined wave packet (localizable particle)
carrying the spin. The second aspect is to characterize the correlations of two spinons in an $S=1$ state, 
to determine whether they are confined, and, if so, to extract the size of the bound state.

\begin{figure}
\centerline{\includegraphics[width=7cm, clip]{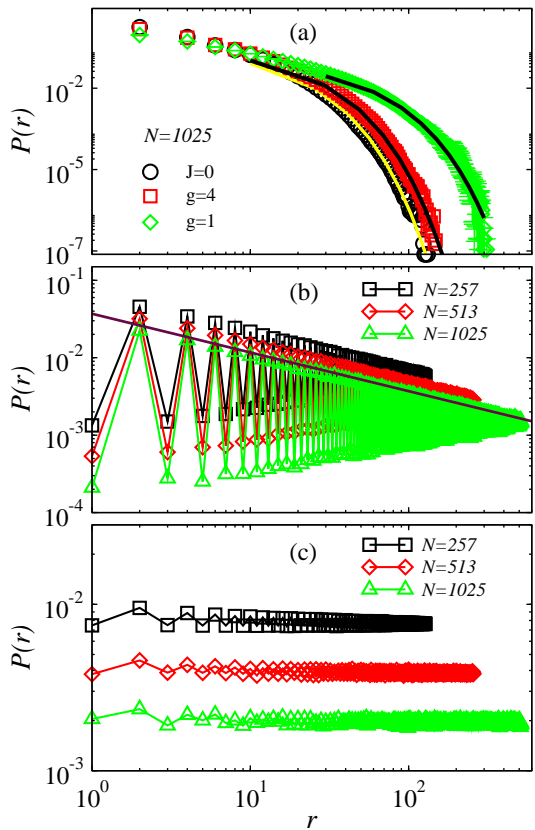}}
\vskip-2mm
\caption{(Color online) Overlap $P(r)=\langle \psi_{1/2}(i+r)|\psi_{1/2}(i)\rangle$ for (a) different VBS states of 
the J-Q$_{\rm 3}$ model of size $N=1025$, (b) at $g_c$ for different $N$, and (c) in the N\'eel state of 
the long-range model ($\alpha=3/2$) for different $N$. The curves in (a) are fits to $\propto {\rm e}^{-r/\lambda}$ 
(with $\lambda \approx 9$ at $J=0$) and the line in (b) shows the form $\propto 1/\sqrt{r}$.} 
\label{fig3}
\vskip-3mm
\end{figure}

The $S=1/2$ ground state (with momentum $k=0$) can be written as $|\Psi_{1/2}\rangle =\sum_r |\psi_{1/2}(r)\rangle$, where $r$ 
is the location of the unpaired spin \cite{banerjee10}. Denoting a basis state with the spinon at $r$ as $|V_\alpha(r)\rangle$, we have
 $|\psi_{1/2}(r)\rangle = \sum_\alpha f_r^{\alpha}|V_\alpha(r)\rangle$ and the overlap of two states with different location of
their spinon cores is 
\begin{equation}
\langle \psi_{1/2}(r')|\psi_{1/2}(r)\rangle =
\sum_{\alpha\beta} f_{r'}^{\beta} f_r^{\alpha}  \langle V_\beta (r')|V_\alpha(r)\rangle.
\label{olap1}
\end{equation}
What we propose is that this quantity allows for a generic way to test whether a spinon is a well
defined particle. Such a particle should have a finite wave packet (i.e., a minimum size of a region 
to which the $S=1/2$ degree of freedom can be confined), which typically should lead to an exponential 
decay of the overlap with the separation $|r'-r|$ (with a power-law decay corresponding to a marginal 
case). This follows in a VBS state because the basis-state overlap $\langle V_\beta(r')|V_\alpha(r)\rangle$ 
is dictated by the number of loops in the transition graph. An $S=1/2$ transition graph has a string of 
bonds terminating in unpaired spins \cite{banerjee10}, as seen in \ref{fig2}(b). In a VBS state, the loops are typically 
short, and the presence of a string will reduce the number of loops in proportion to the length of the string,
and, thus, $\langle V_\beta(r')|V_\alpha(r)\rangle$ and (\ref{olap1}), should decay exponentially with the 
separation $|r'-r|$. One can then also expect a power-law decay in a critical VBS state.

The overlap (\ref{olap1}) can be computed by accumulating the distribution 
$P(r)$ of separations $r$ of the unpaired spins in the $S=1/2$ transition graphs. The above expected behaviors 
are indeed realized in the J-Q$_{\rm 3}$ model, as shown in Fig.~\ref{fig3}. In VBS states for large $N$, the 
overlap vanishes for odd distances, implying that the bra and ket spinons are on the same sublattice in the 
infinite system. For the even distances the overlap is of the form $P(r) \propto {\rm e}^{-r/\lambda}$, and $\lambda$
is essentially the size of an exponentially decaying wave packet. 
%(although not exactly, since the overlap has a correction to the dominant exponential decay). 
The size $\lambda$ is roughly twice the spin correlation length in the cases we have studied.

In the critical state, the overlap has the form $P(r) \sim 1/\sqrt{r}$ and the wave packet is only marginally 
defined. The Heisenberg chain is known to have spinon excitations \cite{faddeev81} and, thus, it appears that one can still 
consider such a broad algebraic wave packet as a particle. The total weight of all odd-$r$ overlaps is roughly constant, 
$\approx 1/4$.

In the N\'eel state $P(r)$ is almost flat and even and odd $r$ have almost the same weight. The unpaired spin in the N\'eel 
state is, thus, not localizable within a wave packet, in agreement with the expectation that the spinon should not be an elementary 
excitation of this state. The unpaired spin is strongly aligned with the N\'eel order of the rest of the system 
\cite{sanyal11} and cannot be regarded as an independent spatial $S=1/2$ degree of freedom. 
%It is interesting that this is manifested so clearly in the overlap (\ref{olap1}).

\begin{figure}
\centerline{\includegraphics[width=6.7cm, clip]{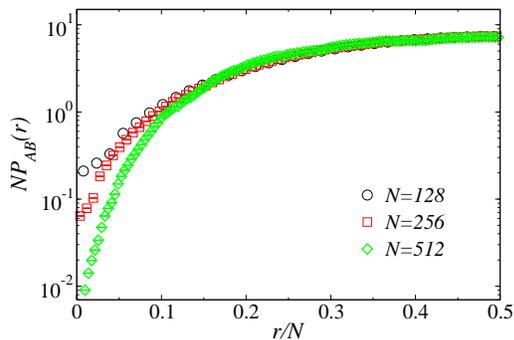}}
\vskip-2mm
\caption{(Color online) Size-scaled distribution of the distance between two spinons in the $S=1$ 
VBS state at $g=4$.}
\label{fig4}
\vskip-3mm
\end{figure}

For an $S=1$ state with two unpaired spins we have
\begin{align}
&\langle \psi_{1}(r_{A}',r_{B}')|\psi_{1}(r_A,r_B)\rangle \nonumber \\ 
&~~~ = \sum_{\alpha\beta} f_{r_{A}'r_{B}'}^{\beta} f_{r_{A}r_{B}}^{\alpha} 
\langle V_\beta (r_{A}',r_{B}')|V_\alpha(r_{A},r_{B})\rangle, 
\label{olap2}
\end{align}
where, as indicated, in both the bra and the ket state one spinon is on sublattice A and one on B. Here we can define 
several probability distributions depending on a single distance, e.g., $|r_A-r_B|$ or $|r_A-r_B'|$, integrating over the 
remaining two free-spin locations. To investigate the confinement length we define $P_{AB}(r)$ as the average of the distributions 
of the above two distances. The results, shown in Fig.~\ref{fig4}, indicate deconfined spinons with weak mutual repulsion, which 
makes the distribution broadly peaked at $r=N/2$. Size-scaled distributions $NP_{AB}(r)$ for different $N$ fall almost on top of 
each other when graphed versus $r/N$. For confined spinons, the confinement length will be reflected in an asymptotic decay 
$P_{AB}(r) \sim {\rm e}^{-r/\Lambda}$, where $\Lambda$ is the size of the bound state.

\begin{figure}
\centerline{\includegraphics[width=6.7cm, clip]{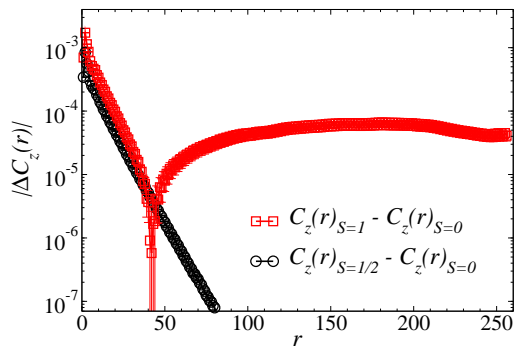}}
\vskip-2mm
\caption{(Color online) Spin correlations in an $N=513$ system with a single spinon ($S=1/2$) and in $N=512$ 
systems with two spinons ($S=1$) in the $J=0$ VBS state. The $N=512$, $S=0$ correlation has been subtracted 
off to isolate the spinon contributions. For $S=1$ there is a phase change at $r \approx 42$.} 
\label{fig5}
\vskip-3mm
\end{figure}

The two length-scales we have discussed---the size of the spinon wave packet $\lambda$  and the bound state $\Lambda$  (for confined 
spinons)---are also visible in the $z$-component spin correlation function $C_z(r)=\langle S^z_{i}S^z_{i+r}\rangle$. We demonstrate this 
for both $S=1/2$ and $S=1$ states in Fig.~\ref{fig5}. The reason for the spinon contributions can be understood from Figs.~\ref{fig2}(b,c). 
The unpaired spins in the $S=1$ state (c) dominate the long-distance correlation function if the confinement length is larger 
than the correlation length of the background VBS. In addition, the short-distance correlations of both the $S=1/2$ and $S=1$ states 
are modified by the presence of strings. Indeed, as we demonstrate in Fig.~\ref{fig5}, by subtracting off $C_z(r)$ of the $S=0$ state, 
the remaining short-distance correlations contain an exponentially decaying contribution which for both $S=1/2$ and $S=1$ is roughly 
twice the correlation length, i.e., similar to the wave packet size $\lambda$. In the $S=1$ state, the correlation function remains
non-zero, $\propto 1/N$, as $r\to \infty$, reflecting deconfined spinons. Note that there is a change in phase of $C_z(r)$, at
some $r$ which is related to $N$ and $\lambda$ (and hence depends on $g$).

{\it Conclusions and discussion}---We have introduced a method to determine whether a spinon is a 
well-defined emergent particle (excitation) of a quantum spin system, and, if so, whether two spinons
in an $S=1$ excitation are deconfined or form a bound state (the size of which can be computed). The 
discussion was framed around the valence-bond basis and QMC simulations with it, but the definitions 
are independent of this basis. Our arguments only rely on the fact that one can write a state for, 
e.g., $S=1/2$ as $\sum_r |\psi_0(r)\rangle \otimes |\uparrow_r\rangle$ (for momentum $k=0$, with 
self-evident generalization to $k\not=0$), where $|\psi_0(r)\rangle$ is an $S=0$ state of all 
spins except the one at $r$ (and similar decompositions for higher $S$). One can, thus, compute the 
quantities we have investigated here with other methods as well. The crucial observation is that states 
$|\psi_0(r)\rangle \otimes |\uparrow_r\rangle$ for different $r$ are non-orthogonal. If the unpaired spin 
$\uparrow_r$ is localized within a spinon wave packet (by definition for a spinon), 
then the overlaps give direct information on the size of this wave packet. The spinon is not an
independent particle if the wave packet is uniformly delocalized over the whole system as $N\to \infty$, as
we have demonstrated here for a N\'eel state. 

Our method does not rely on any knowledge or theory of the nature of the 
spinon (other than it carrying spin $S=1/2$). The wave-function 
overlaps (\ref{olap1}) and (\ref{olap2}) are completely general and applicable 
to any system in any number of dimensions. For 1D systems 
there are alternative methods to study spinons using the fact 
that they are kink and antikink solitons \cite{ssh}, 
which can be created by boundary conditions.
The spinon wave function, which is similar to that of 
a particle in a box \cite{sorensen},
does not, however, contain any direct information 
on the intrinsic size of the spinon ``particle''. 
A criterion of deconfinement based on impurity (un)binding was also presented recently \cite{doretto09}, but that approach cannot 
unambiguously determine whether a spinon is a well defined particle. Our approach also avoids potential differences 
between spinon-spinon and spinon-impurity affinities

%As a natural next step, it will be interesting to determine if spinons are associated 
%with a new length-scale in 2D J-Q models \cite{sandvik07,lou09}, as has been predicted theoretically 
%for 2D N\'eel-VBS transitions \cite{senthil04}.

{\it Acknowledgments}---We would like to thank C. Batista, K. Damle, and S. Shastry for discussions.
This work was supported by NSF Grant No.~DMR-0803510. 

\null

\end{document}